\documentclass[floatfix]{revtex4}

\usepackage{blindtext}

\usepackage{graphicx,epsfig,epstopdf}
\usepackage{amsmath}
\usepackage{color}
\usepackage{subfigure}
\usepackage{array}
\usepackage{tabularx}
\usepackage{multirow}
\usepackage{esvect}
\newcolumntype{L}[1]{>{\raggedright\arraybackslash}p{#1}}
\newcolumntype{C}[1]{>{\centering\arraybackslash}p{#1}}
\newcolumntype{M}[1]{>{\centering\arraybackslash}m{#1}}

\def\_#1{{\bf #1}}
\def\.{\cdot}

\begin{document}

\title{Transfer matrix method for the analysis of space-time modulated media and systems}

\author{
 Junfei Li$^{1,\dag}$, Xiaohui Zhu$^{2,\dag}$, Chen~Shen$^{1}$, Xiuyuan Peng$^{1}$ and Steven~A.~Cummer$^{1}$
}
 
\affiliation{$^1$Department of Electrical and Computer Engineering, Duke University, Durham, North Carolina 27708, USA\\
$^2$ School of Mechatronics Engineering, Harbin Institute of Technology, Harbin, Heilongjiang 150001, China\\
$^{\dag}$ J. L. and X. Z. contributed equally to this paper.
Correspondence and requests for materials should be addressed to S.A.C. (cummer@ee.duke.edu)}

\begin{abstract} 
Space-time modulation adds another powerful degree of freedom to the manipulation of classical wave systems. It opens the door for complex control of wave behavior beyond the reach of stationary systems, such as nonreciprocal wave transport and realization of gain media. Here we generalize the transfer matrix method and use it to create a general framework to solve wave propagation problems in time-varying acoustic, electromagnetic, and electric circuit systems. The proposed method provides a versatile approach for the study of general space-time varying systems, which allows any number of time-modulated elements with arbitrary modulation profile, facilities the investigation of high order modes, and provides an interface between space-time modulated systems and other systems.
\end{abstract}

\maketitle

\section{Introduction}
Wave propagation in systems where the material parameters or structures are varying in both space and time has attracted considerable attention in recent years. Such space-time modulation provides an efficient means to break time-reversal symmetry and has found many applications in the field of integrated circuits (IC), optics, electromagnetics (EM) and acoustics. For example, time-varying transmission lines (TVTL) have been used to create frequency converters, multipliers and non-reciprocal devices such as isolators and circulators \cite{tien1958traveling,tien1958parametric,cullen1958travelling,simon1960action,oliner1961wave,cassedy1963dispersion,cassedy1967dispersion,elachi1972electromagnetic,qin2014nonreciprocal}. Recently, the idea of achieving non-reciprocity through space-time modulation has been applied to modern optical and electromagnetic systems \cite{hafezi2012optomechanically, sounas2014angular,estep2014magnetic,hadad2015space, correas2016nonreciprocal, ruesink2016nonreciprocity,hadad2016breaking,taravati2017mixer, miri2017optical}. For mechanical waves, space-time modulated elastic beams have been proposed to create a directional band gap \cite{trainiti2016non,trainiti2019time}. Space-time modulated mass-spring systems have also been proposed to study directional wave manipulation for elastic waves \cite{nassar2017non,wang2018observation}. In airborne acoustics, frequency converters and parametric amplifiers have been demonstrated in a space-time modulated metamaterial \cite{li2019nonreciprocal}. Acoustic isolators \cite{shen2019nonreciprocal}, circulators \cite{fleury2014sound,fleury2015subwavelength,koutserimpas2019coupled} and topological insulators \cite{fleury2016floquet} have also been demonstrated with temporally modulated resonators.

Theoretical tools available for studying time-varying wave systems generally fall into several categories. In time varying transmission lines \cite{tien1958traveling,tien1958parametric,cullen1958travelling}, the most commonly used approach is to directly solve the coupled differential equations by assuming slowly varying envelope, and then solve the coupled equations for the envelope. However, such approach only considers a small number of interacting waveguide modes and neglects all the higher order modes. Taking these modes into account will result in additional coupled differential equations and make the system more difficult to solve efficiently. Time-Floquet theory \cite{floquet1883equations,trainiti2016non,nassar2017non}, on the other hand, calculates the band structure for an infinitely long system. However, the Floquet theory only predicts which waveguide modes are coupled, and it does not provide detailed information on how do waves change gradually in such systems. Both theoretical approaches deal with infinitely long and continuous systems. While in practice, the systems must have a finite length and, in many cases, the realization for space-time modulated media are discretized \cite{qin2014nonreciprocal,wang2018observation,trainiti2019time,li2019nonreciprocal}. Furthermore, both solving coupled wave equations directly and time-Floquet theory deal with wave propagation where the material properties are modulated sinusoidally and they cannot be applied to more complicated modulation profiles. For discrete systems with space-time varying boundary conditions, the system can be solved by balancing the harmonics at each order \cite{hadad2015space,shen2019nonreciprocal}. Another type of discrete system involves one or more coupled resonators and is generally solved by the coupled mode theory (CMT) \cite{yariv1973coupled,haus1991coupled}. CMT has been widely applied in mechanical-optical systems  \cite{li2010coupled,ruesink2016nonreciprocity,miri2017optical} and is recently introduced in acoustics \cite{fleury2015subwavelength,koutserimpas2019coupled}. However, the coupling coefficients are usually not easy to determine or design in practice. Furthermore, for both harmonic balancing and CMT, as the number of resonators or boundaries increase, and with the increase of modes taken into account, the coupled equations becomes complicated to solve.

Here we propose a generalized transfer matrix approach for solving space-time varying systems. By setting up time-varying boundary conditions and rewriting them into transfer matrix form, the effects of time variation are localized. Therefore, all the time-varying components can be described individually so that we can calculate a system with arbitrary number of time-varying elements and arbitrary modulation profile by simply multiplying their transfer matrices. It is shown that with small number of elements, it reduces to harmonic balancing method described in \cite{shen2019nonreciprocal}, while with large number of modulated elements with small spacing, it reduces to the continuous counterpart \cite{li2019nonreciprocal} and time Floquet theory. The results are verified with finite-difference time-domain (FDTD) simulations. Our approach provides a versatile platform to investigate the behavior of general space-time modulated systems. Compared with current available theoretical tools, it has a number of advantages. First, it handles multiple higher-order waves and arbitrary number of modulated elements without increasing the computational complexity. Second, it offers many degrees of freedom and thus allows the study of arbitrary modulation parameters such as modulation depth and phase for each element, or even the study of a system with random time modulation, which is difficult for other approaches. Third, it has no constraint for the element spacing or material properties. Therefore, it facilities the study of interplay of time-dependent wave behavior with other classical behaviors, such as resonances, multiple scattering and inhomogeneity. Fourth, by representing the whole system with a transfer matrix, our method provides an interface for the study of interaction between a space-time modulated system with other stationary and non-stationary systems. 

The paper is organized as follows. First, we present the formulation of transfer matrix method for general time-varying systems. We used acoustic representation and the results can be directly applied to EM waves and ICs, and the corresponding formulation can be found in the APPENDIX A. The approach is then applied to several examples to show its capabilities and advantages. In the first case, we present the design of an acoustic diode without operating at resonance frequencies. The isolation level can be controlled by employing different number of modulated resonators and modulation strategies. The off-resonance feature makes the design robust to loss and fabrication errors. In the second case, we will show that when the phase-matching condition is met, parametric mode conversion and amplification can be achieved with multiple space-time modulated resonators. With this example, we will discuss the impact of high-order modes, multiple scattering, and how this approach reduces to continuous theory and time Floquet theory under several conditions.

\section{Matrix representation of a space-time varying system}
In this section, we exemplify the derivation with the acoustic representation. The general procedure applies to the study of EM waves and ICs, where one just need to substitute the pressure and velocity $(p,v)$ with electric and magnetic field $(E,H)$ or voltage and current $(V,I)$. The details for the EM and IC formulation, and the corresponding realization approaches are summarized in APPENDIX A.
\subsection{Response of time varying load in series}
Consider an acoustic waveguide loaded with a time-varying impedance sheet $Z_L$. Assume the impedance is varying harmonically in the form of 
\begin{equation}
    Z_L(\omega,t)=Z_{L0}(\omega)[1+a\cos{(\Omega t+\phi)}],
\end{equation}
where $\Omega$ is the modulation frequency, $\phi$ is the initial phase of the modulation, $a$ denotes the modulation depth, and $Z_{L0}$ is the impedance without modulation. A plane wave is launched into the waveguide with angular frequency $\omega_0$. Due to the time-varying load impedance, harmonics will be generated. Therefore, the pressure and velocity on the upstream and downstream of the load are written as
\begin{gather}
    p_{\mp}=\sum_{n=-\infty}^{\infty}p_{\mp}^ne^{j\omega_nt}\\
    v_{\mp}=\sum_{n=-\infty}^{\infty}v_{\mp}^ne^{j\omega_nt},
\end{gather}
where $\omega_n=\omega_0 \pm n\Omega$. The boundary condition at the position of the load should satisfy
\begin{gather}
    p_--p_+=v_+ Z_L\label{eq:BC_p1}\\
    v_-=v_+,
    \label{eq:BC_v1}
\end{gather}

Put the expression of the series load, pressure and velocity into Eqs. (\ref{eq:BC_p1}) and (\ref{eq:BC_v1}), equate the terms with $e^{j\omega_nt}$ using the relation $\cos(\Omega t+\phi)=\frac{1}{2}[e^{j(\Omega t+\phi)}+e^{-j(\Omega t+\phi)}]$, we can rewrite the boundary conditions in terms of each order of the harmonics:
\begin{gather}
    \begin{aligned}
    p_-^n-p_+^n=Z_{L0}^n v_+^n+\frac{a Z_{L0}^{n-1}}{2}e^{j\phi}v_+^{n-1}\\
    +\frac{a Z_{L0}^{n+1}}{2}e^{-j\phi} v_+^{n+1}
    \end{aligned}\\
    v_-^n=v_+^n
\end{gather}
where the superscript denotes the impedance at each order of harmonic, i.e., $Z_{L0}^{n}=Z_{L0}(\omega_n)$ and so on. The transfer matrix $M$ is defined as
\begin{equation}
    \begin{bmatrix}
	\vdots\\
	p_-^{n-1}\\
	v_-^{n-1}\\
	p_-^{n}\\
	v_-^{n}\\
	p_-^{n+1}\\
	v_-^{n+1}\\
	\vdots
	\end{bmatrix}=M    \begin{bmatrix}
	\vdots\\
	p_+^{n-1}\\
	v_+^{n-1}\\
	p_+^{n}\\
	v_+^{n}\\
	p_+^{n+1}\\
	v_+^{n+1}\\
	\vdots
	\end{bmatrix}.
\end{equation}
With the boundary conditions, the transfer matrix at the load can be written as
\begin{equation}
M=    \begin{bmatrix}
	  & \vdots & \vdots & \vdots & \vdots & \vdots & \vdots & \\
	\dots & 1 & Z_{L0}^{n-1} & 0 & \frac{a Z_{L0}^{n}}{2}e^{-j\phi} & 0 & 0 & \dots\\
	\dots & 0 & 1 & 0 & 0 & 0 & 0 & \dots\\
	\dots & 0 & \frac{a Z_{L0}^{n-1}}{2}e^{j\phi} & 1 & Z_{L0}^{n} & 0 &\frac{a Z_{L0}^{n+1}}{2}e^{-j\phi} & \dots\\
	\dots & 0 & 0 & 0 & 1 & 0 & 0 & \dots\\
	\dots & 0 & 0 & 0 & \frac{a Z_{L0}^{n}}{2}e^{j\phi} & 1 & Z_{L0}^{n+1}& \dots\\
	\dots & 0 & 0 & 0 & 0 & 0 & 1 & \dots\\
	 & \vdots & \vdots & \vdots & \vdots & \vdots & \vdots & 
	\end{bmatrix}\label{eq:Mseries}
\end{equation}

For waves travelling in an empty waveguide, the transfer matrix can be simply written as
\begin{equation}
    M_T=
    \begin{bmatrix}
	 & \vdots & \vdots & \vdots & \\
	\dots & M_T^{n-1} & 0 & 0 & \dots \\
	\dots & 0 & M_T^{n} & 0 & \dots \\
	\dots & 0 & 0 & M_T^{n+1} & \dots \\
	 & \vdots & \vdots & \vdots &  
	\end{bmatrix}
\end{equation}

where

\begin{equation}
    M_T^i=
    \begin{bmatrix}
    \cos(k_i d) & jZ_0 \sin(k_i d)\\
	\frac{j}{Z_0} \sin(k_i d) & \cos(k_i d)\\
    \end{bmatrix}, (i=...n-1, n, n+1...)
\end{equation}
Here $Z_0$ is the characteristic impedance of air, $k_i$ is the wavenumber for the $i$ th order wave, and $d$ is the length of the waveguide. All the coupling terms between each order are 0 since the waveguide is stationary.

For a system composed of $N$ equally-spaced cascaded time-varying impedance sheets, the transfer matrix can be calculated as $M=M_1 M_T M_2 M_T \dots M_T M_N$, where $M_i(i=1,\dots,N)$ denotes the transfer matrix of the impedance loads, and $M_T$ is the transfer matrix of each section of the empty waveguide.

\subsection{Response of time varying load in parallel}

For a shunted load on the waveguide, it is more convenient to use the effective admittance, defined as $Y_L=S_L/(S_w Z_L)$ where $S_w$ and $S_L$ denote the cross sectional area of the waveguide and the load, respectively. Note here that the admittance is scaled by the ratio between the cross sectional areas to keep the generality of our formulation. In this case, the boundary condition at the position of the load should satisfy
\begin{gather}
    p_-=p_+\label{eq:BC_p2}\\
    v_--v_+=Y_L p_+,
    \label{eq:BC_v2}
\end{gather}
 Now if we assume the admittance varies in the form of $Y_L(\omega,t)=Y_{L0}(\omega)[1+a\cos{(\Omega t+\phi)}]$, the boundary conditions for each order then become

\begin{gather}
    p_-^n=p_+^n\\
    \begin{aligned}
    v_-^n-v_+^n=Y_{L0}^n p_+^n+\frac{a Y_{L0}^{n-1}}{2}e^{j\phi}p_+^{n-1}\\
    +\frac{a Y_{L0}^{n+1}}{2}e^{-j\phi}p_+^{n+1}.
    \end{aligned}
\end{gather}

In this case, the transfer matrix at the load becomes
\begin{widetext}
\begin{equation}
M=    \begin{bmatrix}
	  & \vdots & \vdots & \vdots & \vdots & \vdots & \vdots & \\
	\dots & 1 & 0 & 0 & 0 & 0 & 0 & \dots\\
	\dots & Y_{L0}^{n-1} & 1 & \frac{a Y_{L0}^{n}}{2}e^{-j\phi} & 0 & 0 & 0 & \dots\\
	\dots & 0 & 0 & 1 & 0 & 0 & 0 & \dots\\
	\dots & \frac{a Y_{L0}^{n-1}}{2}e^{j\phi} & 0 & Y_{L0}^{n} & 1 & \frac{a Y_{L0}^{n+1}}{2}e^{-j\phi} & 0 & \dots\\
	\dots & 0 & 0 & 0 & 0 & 1 & 0 & \dots\\
	\dots & 0 & 0 & \frac{a Y_{L0}^{n}}{2}e^{j\phi} & 0 & Y_{L0}^{n+1} & 1 & \dots\\
	 & \vdots & \vdots & \vdots & \vdots & \vdots & \vdots & 
	\end{bmatrix}\label{eq:Mparallel}
\end{equation}
\end{widetext}

We can see from Eq.~(\ref{eq:Mseries}) and Eq.~(\ref{eq:Mparallel}) that the loads in series and in parallel are similar, which is consistent with the passive cases \cite{pozar2009microwave}. With the transfer matrix for the loads and empty waveguide, the transfer matrix for the entire structure can be calculated by multiplying the transfer matrices for all components. Hence, the transmission and reflection coefficients can be calculated by converting the transfer matrix into scattering matrix. The conversion from transfer matrix to scattering matrix is given in APPENDIX B.

With the transfer matrix method we can, in principle, take all the orders of harmonics into account. However, in practice, the matrix shall be truncated to account only for the orders that are non-negligible. Compared with existing theories that characterize space-time modulated systems, there are three main advantages for the proposed method. First, our proposed transfer matrix method takes higher order modes into account. It will be shown in the following sections that these high-order modes exhibit non-negligible effects on the wave propagation. Second, current theories focus on the travelling wave-like modulation where the modulation has linear phase gradient along the space, while with the proposed theory, the transfer matrix for all the time-varying elements can be written independently. Therefore, by cascading the transfer matrices, we will be able to calculate arbitrary space-time modulation profile. Third, the generalized transfer matrix method provides an interface between time-varying systems and other systems, so it can be used to study the interaction between stationary and non-stationary systems.

\subsection{An acoustic case: waveguide loaded with space-time modulated Helmholtz resonators}
\begin{figure}
	\includegraphics[width=0.6\linewidth]{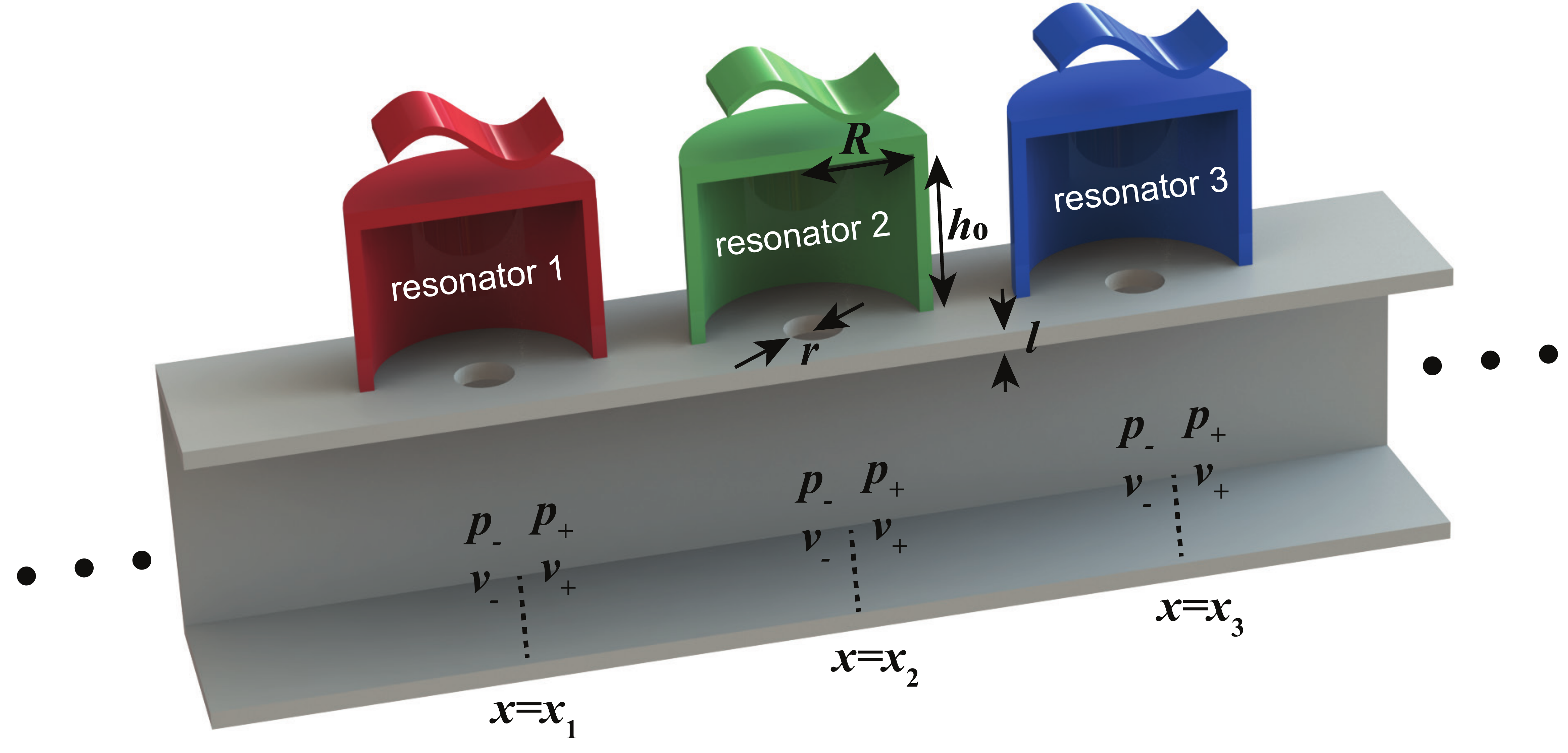}
	\caption{Schematic representation of a waveguide loaded with arbitrary number of space-time modulated resonators}  \label{fig:Resonator_series}
\end{figure}

In this section, we will look into a practical system in acoustics where a waveguide is side-loaded with a series of Helmholtz resonators whose cavity heights are modulated in both space and time. The schematic diagram is shown in Fig. \ref{fig:Resonator_series}. The impedance of a Helmholtz resonator can be represented with $Z=j\omega L+\frac{1}{j\omega C}$ where $L=\rho l$, $C=\frac{S_{cav}h}{S_L\rho c^2}$. Here $\rho$ and $c$ are the density and sound speed in air, $l$ is the corrected length of the neck, $S_L$ and $S_{cav}$ are the area of the neck and cavity, respectively.

Assume the height of the cavity $h$ varies in the form of $h(t)=h_0[1+m\cos(\Omega t+\phi)]$, and the modulation is weak, i.e., $m\ll1$. The impedance of a single Helmholtz resonator can be written as 
\begin{equation}
    Z=Z_n[1-\frac{m Z_{cn}}{Z_n} \cos(\Omega t+\phi)]
    \label{eq:imped_approx}
\end{equation}

where $Z_n=j\omega_n L+\frac{1}{j\omega_n C_0}$ and $Z_{cn}=\frac{1}{j\omega_n C_0}$, $C_0=\frac{S_{cav}h_0}{S_L\rho c^2}$ is the acoustic capacitance for a resonator with no modulation. The impedance variation corresponds to the parallel load case described in Sec. II. Inserting the impedance into the above formalism, a physical acoustic system can be represented by the transfer matrix framework described above.

\section{Examples in the discretized space-time modulated systems}
\subsection{Non-reciprocal sound transport}
 Non-reciprocal devices that create directional control of the energy flow have numerous applications and, consequently, have attracted significant attention in recent years. Non-reciprocal sound transport using space-time modulation has recently been reported by cascading modulated resonators \cite{shen2019nonreciprocal} or forming a circulator \cite{fleury2015subwavelength,fleury2016floquet}. However, the proposed designs require resonators with high quality factor, and the system has to work close to the resonant frequency. These features bring challenges in realization, and make the system sensitive to inevitable losses and fabrication errors. In this section, we will demonstrate the design of non-reciprocal device by cascading multiple space-time modulated resonators. Different from the existing strategy who derives the requirements for the resonators, we start from a physically realizable resonator design, and then determine the number of resonators needed to generate sufficient non-reciprocal response at off-resonance frequencies.
 
 Without loss of generality, we pick an acoustic waveguide with cross section $9.5\rm{mm}\times9.5\rm{mm}$. The Helmholtz resonator dimension is shown in Fig. \ref{fig:Resonator_series}. The cavity is cylindrical with radius $R=14$mm and height $h_0=10$mm. The neck is also cylindrical with radius $r=4.5$mm and effective neck length $l=4.7$mm, yielding a resonance frequency of 2566Hz. The distance between adjacent shunted resonators is 40mm. The modulation frequency is 100 Hz and the modulation depth $m$ is chosen as 0.15. The modulation phase of each resonator has a linear gradient, i.e., the modulation phase for the $n$th resonator can be written as $\phi_n=n\Delta\phi$ where $\Delta\phi$ is the phase difference between adjacent resonators. The incident wave is assumed to be sinusoidal with angular frequency $\omega_0$, and is defined as the 0th order. We have studied the same system truncated at different orders, and found that for this particular system, the transmission coefficients converge after the $\pm 5$th order is considered. The detailed information is summarized in APPENDIX C. In the study, the waves are truncated to $\pm 10$th order.
 
\begin{figure}
	\includegraphics[width=0.8\linewidth]{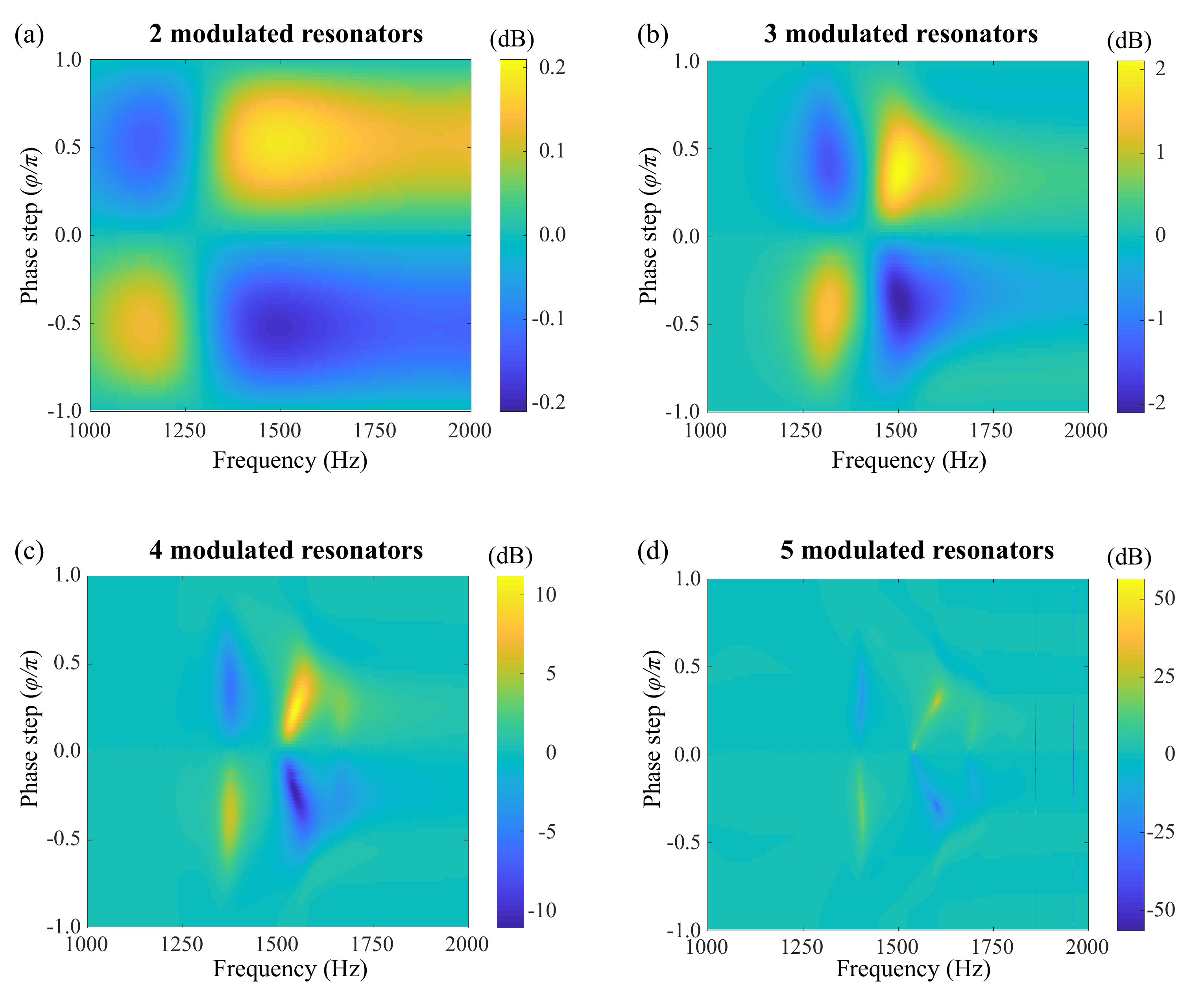}
	\caption{Transmission ratio for 2, 3, 4 and 5 space-time modulated resonators with the change of input frequency and modulation phase step. With the increased number of resonators, the range of available transmission ratio can be increased.}  \label{fig:adding_resonators}
\end{figure}
 
 By multiplying transfer matrices and converting the total transfer matrix into scattering matrix, the transmission and reflection coefficients for such a system can be calculated. 
 Since we truncated the waves to $\pm 10$th order, the size of scattering matrix is 42 by 42, and $S_{22,21}$ and $S_{21,22}$ represent the transmission coefficients for the 0th order wave in the forward direction and backward direction (see APPENDIX B for details). Here we define the transmission ratio $\Gamma=|S_{22,21}|/|S_{21,22}|$ as a measure of resulting asymmetry. In Fig. \ref{fig:adding_resonators}, we show the change of $\Gamma$ by varying the phase difference $\Delta\phi$ from 0 to $2\pi$ and incident frequency from 1000Hz to 2000Hz, in the cases of 2, 3, 4 and 5 cascaded resonators.
 
 With two modulated resonators, the asymmetric modulation creates a directional bias, which leads to non-reciprocity. However, due to the low quality factor of the resonance and that the incident wave is not close to the resonant frequency, the non-reciprocal effect is very weak. In this case, the maximum value of $\Gamma$ is 1.022 (0.189dB). As the number of resonators increases, the maximum transmission ratio reaches 1.274 (2.103dB), 3.585 (11.089dB) and 51.55 (34.245dB) with 3, 4 and 5 modulated resonators. This indicates that with more resonators which introduce more design degrees of freedom, nonreciprocal transmission can be realized conveniently using this approach.

\begin{figure}
	\includegraphics[width=0.8\linewidth]{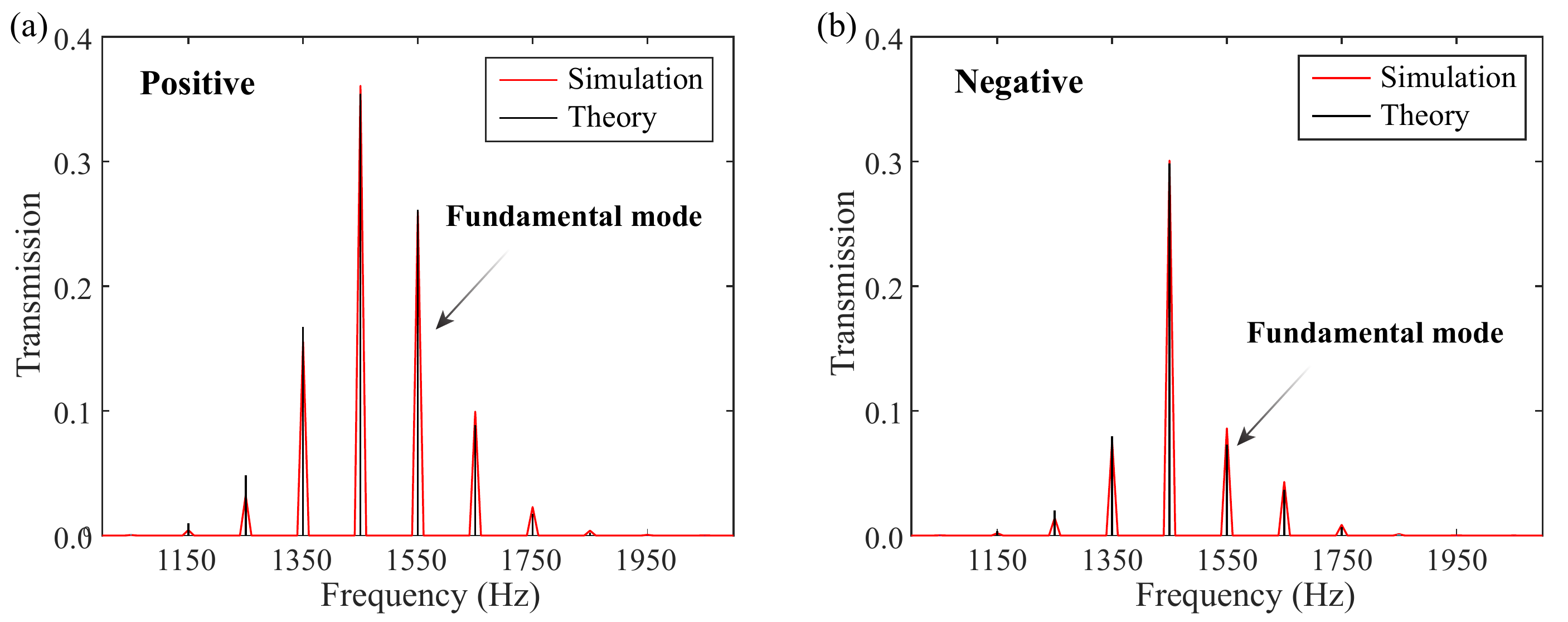}
	\caption{Comparison of the transmission coefficients for each order between theoretical calculation and FDTD simulation. The amplitudes of all the involved orders agree well. Non-reciprocal sound transport is achieved at the fundamental frequency of 1550 Hz.}  \label{fig:Nonrec_comparison}
\end{figure}

The analytic model is verified with 1D FDTD simulation. The background media is lossless air with density $\rho_0=1.21\rm{kg/m^3}$ and speed of sound $c_0=343\rm{m/s}$. The time step is $1\times10^{-7}$s and the grid is $5\times10^{-5}$m. Here we use the 4-resonator case as an example. The dimensions for the Helmholtz resonator are the same as we used in the analytic calculation. In the simulation, each of the 3D resonator is represented with a time-dependent harmonic oscillator and they couple into the 1D waveguide by inducing the discontinuity in local velocity at each position. As an example, the phase step for each resonator and the incident frequency are chosen as $\Delta\phi=0.24\pi$ and 1550 Hz, respectively. The waveguide before and behind the structure is 0.5m, and radiation boundaries are applied to eliminate reflection from both sides. Fig. \ref{fig:Nonrec_comparison} shows the comparison between theoretical calculated transmission coefficients and the simulation results. We can see excellent agreement between them where not only the fundamental order, but also all the higher order waves are well captured. The results again confirm that for multiple resonators, higher-order harmonics need to be considered to accurately represent the system. The 0th order transmission coefficients for the positive direction are 0.2612 and 0.2598 in calculation and simulation, respectively, and for the negative direction, they are 0.0729 and 0.0860. The small discrepancy may come from the first-order truncation in Eq.~(\ref{eq:imped_approx}) for impedance approximation.
 
\subsection{Parametric frequency conversion and parametric amplification}
It has been shown that in TVTL, parametric frequency conversion and amplification can be achieved if the phase-matching condition is met \cite{tien1958traveling,tien1958parametric,qin2014nonreciprocal}. Similar phenomena have also been demonstrated in acousto-optic effects \cite{ruesink2016nonreciprocity,miri2017optical} and space-time modulated media \cite{li2019nonreciprocal}. However, the realization of space-time modulated effective material usually means approximating a continuous system. In many practical cases, however, it is difficult to discretize the system into deep sub-wavelength scales. In these cases, multiple scattering may have significant impact on the wave behavior. Therefore, our theoretical model have great advantage in characterizing such discretized space-time varying systems since the scattering is intrinsically built into the model. In this section, we will show that parametric frequency conversion and parametric amplification can be achieved in such a discrete space-time modulated system. It will also be shown that as we truncate the model to only considering the fundamental mode and targeted mode, the results agree well with those obtained with TVTL and space-time modulated metamaterials. On the other hand, as we will discuss in this section, that in many cases, high-order modes and multiple reflections have non-negligible effects on the fundamental modes and therefore, neglecting them will not yield accurate predictions.

The system configuration we use in this section is as follows: the cross section of the acoustic waveguide is $20\rm{mm}\times20\rm{mm}$; the cavity of the Helmholtz resonator is cylindrical with radius $R=10$mm and height $h_0=5$mm; the neck is also cylindrical with radius $r=1.5$mm and effective neck length $l=3.1$mm. These parameters yield a resonance frequency of 2091 Hz. Note here that we use different set of parameters than the previous case so that the influence of high-order modes and discretization can be better visualized. The distance between adjacent shunted resonators is 40mm. The modulation depth $m$ is chosen as 0.15. The effective dispersion curve can be obtained in simulation by effective parameter retrieval \cite{smith2005electromagnetic}. The simulation is done with commercial finite element analysis package COMSOL Multiphysics.

\begin{figure}
	\includegraphics[width=0.8\linewidth]{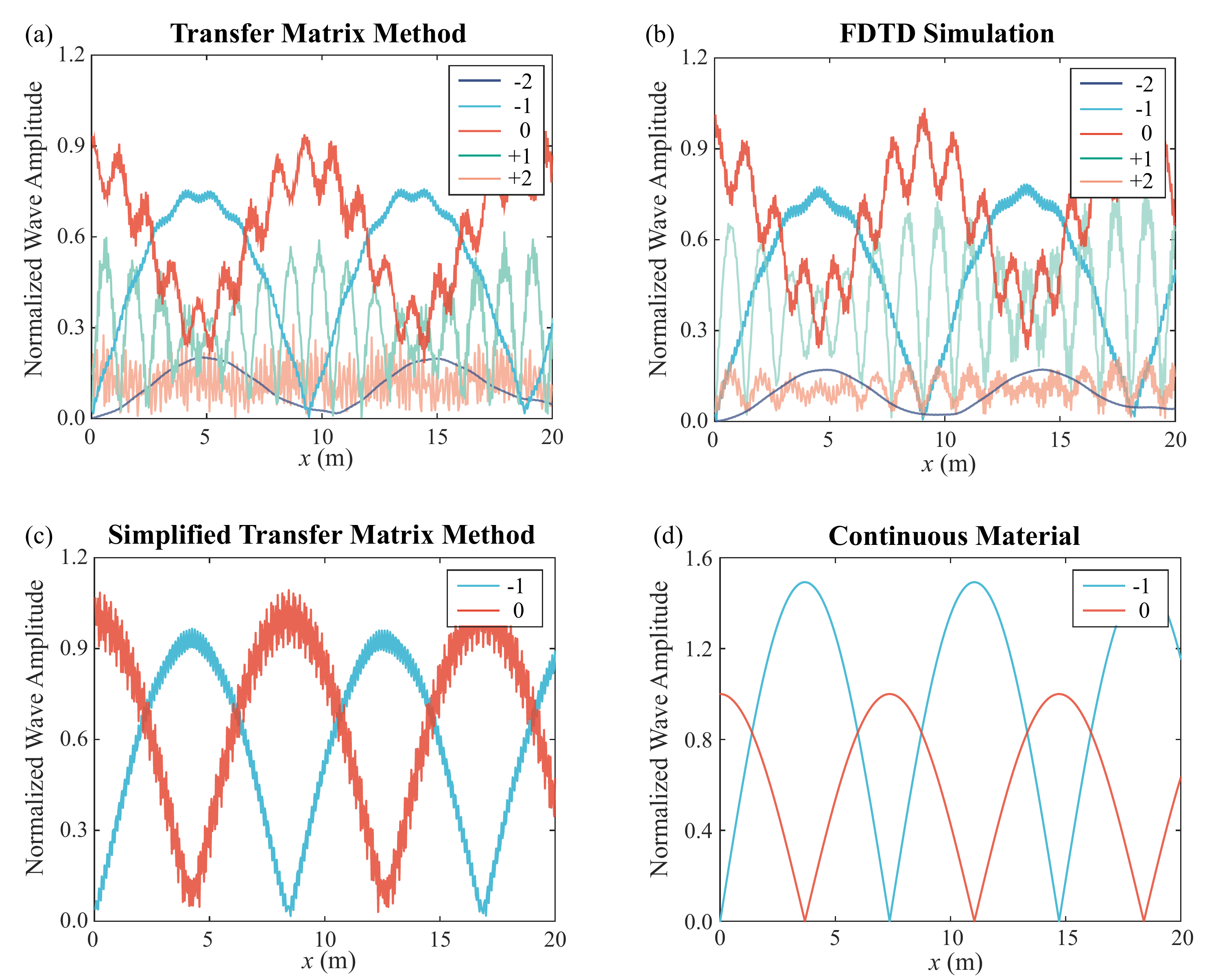}
	\caption{Comparison between theory and simulation for the case of parametric frequency conversion. (a) The wave amplitudes for 0 to $\pm5$th order waves along the waveguide predicted by the transfer matrix method. The calculation truncated at $\pm10$th order. (b) Amplitudes for 0 to $\pm5$th order waves obtained from simulation. (c) Wave amplitudes calculated by the transfer matrix method when only 0th and -1st order is considered, showing that high-order waves have non-negligible effect. (d) The corresponding calculation using the continuous model as a reference.}  \label{fig:freq_conv}
\end{figure}

The first case we show in this section is the realization of parametric frequency conversion. The total length of the modulated section is 20 m, containing 500 space-time modulated resonators. Two empty waveguides are connected to both sides of the modulated section. In this case, we use 1600 Hz wave as the fundamental mode (0th order) and 1300 Hz wave as the target mode (-1st order). The wave numbers for the two modes are retrieved as $k_1=32.69$ rad/m and $k_2=25.67$ rad/m. Therefore, the modulation frequency for each resonator is 300 Hz and the phase step is $\Delta \phi=(k_1-k_2)\Delta d=0.28$ rad. Assuming a monochromatic wave (0th order) is incident from the empty waveguide. Figure \ref{fig:freq_conv}(a) shows the variation of normalized amplitudes of each mode as the wave propagates along the system. Here the wave amplitudes are obtained by calculating the transmission and reflection coefficients of the space-time modulated system. This is done by computing the total transfer matrix and converting it to scattering matrix by assuming air on both sides. Hence, the pressure and velocity for each mode at the incident port can be calculated. By doing this, the interaction between the space-time modulated system and two empty waveguides can be conveniently analyzed. With total pressure and velocity on the input side, the pressure amplitudes for each order, normalized by the incident wave, at an arbitrary position can be calculated with the help of transfer matrix. FDTD simulation is performed to verify the theoretical results. In the simulation, the time step and space step are $1\times 10^{-7}$s and $5\times 10^{-5}$m, respectively. Radiation boundaries are applied on both ends of the empty waveguides to eliminate reflection. Sinusoidal wave is incident from the air section, and we wait 0.4 seconds for the system to reach its steady state. The waveform along the modulated section from 0.3s to 0.4s is recorded and Fourier Transform is performed to analyze its spectrum. Figure. \ref{fig:freq_conv}(b) shows how energy transfers back and forth among each modes while propagating in the system. The simulation results are in good agreement with the theoretical predictions. The theoretically calculated distance needed for one cycle of such transition is 9.426 m, while in simulation, the measured distance is 9.069 m. We can see that apart from the incident mode and the target mode, the $+1$st mode and $\pm2$nd modes all have non-negligible amplitudes. The small oscillation in the amplitudes is a result of interaction with higher order modes. This can be confirmed by truncating the model to only account for the fundamental mode and the target mode, with all other modes neglected. The mode amplitudes calculated from the simplified model are shown in Fig. \ref{fig:freq_conv}(c) and we can see that the small oscillation disappears when the high order modes are turned off. This confirms our conclusion that high-order modes cannot be neglected while treating the time-varying systems. As a comparison, the corresponding results calculated from a continuous effective material model\cite{li2019nonreciprocal} is shown in Fig. \ref{fig:freq_conv}(d). We can see the energy swapping between two modes, but the amplitudes of two modes are not accurate and the prediction of the distance needed for a cycle of transition is $7.357$m, which is $18.9\%$ off from the simulation results. This is because in the continuous model, plane wave solutions are assumed. Therefore, with the effective medium approximation, the unit cells are required to be subwavelength. However, this assumption no longer holds near the resonant frequency range, as the wave velocity becomes very slow such that the wavelength is comparable to the spacing between resonators. In our case, the wavelength for the $-1$st, $0$th and $1$st modes are $24.5$cm, $19.2$cm and $14.1$cm, respectively. They are comparable to the separation of $4$cm and don't satisfy the subwavelength condition. Therefore, to better predict the wave behavior, the Bragg scattering needs to be considered.

\begin{figure}
	\includegraphics[width=0.8\linewidth]{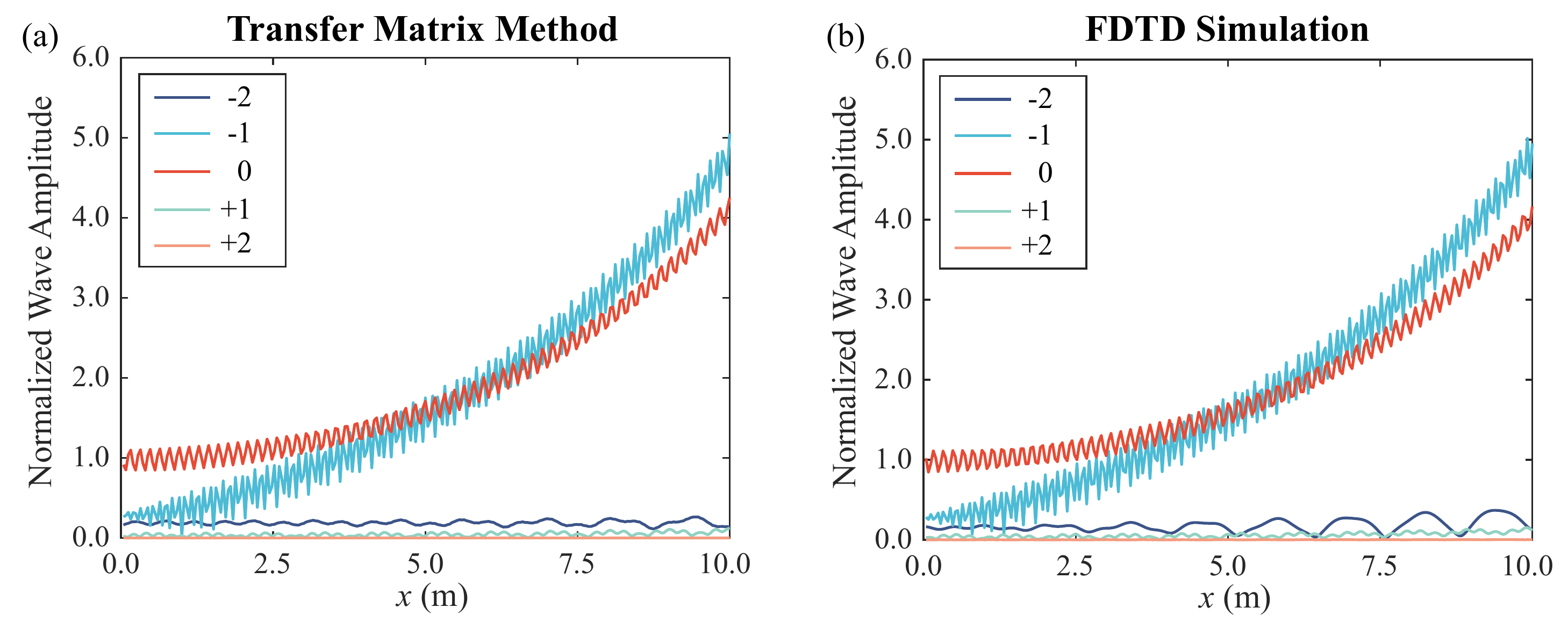}
	\caption{Comparison between theory and simulation for the case of parametric amplification (non-Hermitian gain media). (a) The wave amplitudes for 0 to $\pm2$nd order waves along the waveguide predicted by the transfer matrix method. The calculation truncated at $\pm10$th order. (b) Amplitudes for 0 to $\pm2$th order waves obtained from simulation.}  \label{fig:para_amp}
\end{figure}

The same resonator geometry is used for the parametric amplification case. To better show how the waves grow gradually, the length of modulated section is shortened to $10$m. In this case, the modulation frequency is set as 2500 Hz. The frequencies for the incident wave and generated wave are $1000$ and $1500$Hz, corresponding to wave numbers $k_1=19.46$rad/m and $k_2=30.18$rad/m, respectively. Therefore, the phase step between each resonator is $\Delta \phi=(k_1+k_2)\Delta d=1.99$rad. The comparison between theory and simulation is given in Fig. \ref{fig:para_amp}, where exponential growth for both waves are observed, following the form of $\cosh (\alpha x)$ for $1000$Hz wave and $\sinh (\alpha x)$ for $1500$Hz wave, where $\alpha$ represents the growth rate. $\alpha$ is obtained by fitting the data to the function, and is found to be $0.2056$rad/m and $0.2036$rad/m for theory and simulation, showing excellent agreement. The small variation is the result of discretization. In both theoretical calculation and simulation, the high-order waves are small because these modes are far from the allowed modes in the system, so that the coupling is very weak.

\section{Discussion and Conclusion}
In this paper, we proposed a generic framework to solve time-varying wave propagation problems in acoustic, EM and IC systems through a generalization of the transfer matrix method. By analyzing the boundary conditions for time-varying impedance in series and in parallel, the boundary conditions can be set up by balancing the harmonics. The boundary condition is converted to a generalized transfer matrix form which facilitates the computation process. With this step, the influence of a time-varying element to the whole system is completely localized. Localization provides the advantage of making the time-varying components to serve as individual building blocks, and thus allows the study of arbitrarily complicated time-varying system by simply multiplying individual pieces.

Compared with currently available models for studying general space-time varying systems, our approach has many advantages. First, it enables the study of arbitrary modulation strategies instead of simple sinusoidal travelling wave-like modulation. While dealing with travelling wave-like system, it can reduce to the continuous counterpart, such as solving the coupled partial differential equations (as used in TVTL and space-time modulated metamaterials) and time Floquet theory, with much more precise details. Second, high-order modes are taken into consideration, and we showed the influence of high-order harmonics with two examples: designing non-reciprocal sound transporting device and parametric mode conversion. Third, the computational complexity does not grow with the system complexity, therefore, it is effective in studying large systems, while produces the same results as \cite{shen2019nonreciprocal} when the size of the system is small. Fourth, it allows the study of the interaction between space-time varying behavior and other classical wave behaviors, such as multiple scattering. Fifth, it enables the study of the interaction between a time-varying system and other existing systems, and thus offers possibilities for more advanced and complex wave control.

Space-time modulation puts a new twist to controlling wave behaviors, and opens the door for unprecedented wave manipulation capabilities that are not possible for stationary systems. The possibility offered by modulation is much richer than nonreciprocal wave transport, frequency conversion and the realization of gain media. We hope this theory will serve as a powerful tool and versatile platform for studying the general space-time modulated systems.

\section*{Acknowledgments}
This work was supported by a Multidisciplinary University Research Initiative grant from the Office of Naval Research (Grant No. N00014-13-1-0631) and an Emerging Frontiers in Research and Innovation grant from the National Science Foundation (Grant No. 1641084).

\section*{APPENDIX A: ELECTROMAGNETIC AND IC FORMULATION OF THE SPACE-TIME MODULATED SYSTEMS}
	\setcounter{equation}{0}
	\renewcommand{\theequation}{A\arabic{equation}}
Here we use TE-polarized EM waves as an example. Assume the $\vv{E}$ field is pointing in the $\hat{x}$ direction and the wave is propagating along $\hat{z}$ axis. The waves on the upstream and downstream of a thin impedance sheet connected in series with the waveguide can be written as
\begin{gather}
    \vv{E}_{\mp}=\sum_{n=-\infty}^{\infty}\vv{E}_{\mp}^ne^{j\omega_nt}\\
    \vv{H}_{\mp}=\sum_{n=-\infty}^{\infty}\vv{H}_{\mp}^ne^{j\omega_nt},
\end{gather}
If the impedance sheet only have an electric response and no magnetic response, the boundary conditions for electromagnetic waves can be expressed as
\begin{gather}
    \vv{E}_-=\vv{E}_+ \label{eq:bc_E1}\\
    \hat{z}\times(\vv{H}_+-\vv{H}_-)=\vv{E}_-Y_L.
    \label{eq:bc_H1}
\end{gather}
where $Y_L$ is the admittance of the sheet. Now if we assume the admittance varies in the form of $Y_L(\omega)=Y_{L0}(\omega)[1+a\cos{(\Omega t+\phi)}]$, the boundary conditions for each order then become

\begin{gather}
    \vv{E}_-^n=\vv{E}_+^n\\
    \begin{aligned}
    (-\hat{z}\times\vv{H}_-^n)-(-\hat{z}\times\vv{H}_+^n)=Y_{L0}^n \vv{E}_-^n+\frac{a Y_{L0}^{n-1}}{2}e^{j\phi}\vv{E}_-^{n-1}\\
    +\frac{a Y_{L0}^{n+1}}{2}e^{-j\phi}\vv{E}_-^{n+1}.
    \end{aligned}
\end{gather}

Comparing Eqs. (\ref{eq:bc_E1},\ref{eq:bc_H1}) and Eqs. (\ref{eq:BC_p2},\ref{eq:BC_v2}), we can see that they take the same form, where $\vv{E}$ corresponds to $p$ and $-\hat{z}\times\vv{H}$ corresponds to $v$. This corresponds to the parallel impedance case in acoustics. By converting these equations into the transfer matrix form, all the field terms will be isolated so that the transfer matrix is exactly the same as we derived in Eq. (\ref{eq:Mparallel}). Similar equations can be found for an impedance sheet with time-varying magnetic responses.

For IC systems, the analogy is straightforward. By replacing $p$ with $V$ and $v$ with $I$, all the matrices can be obtained and remain the same form as in the acoustic formulation. Here $V$ and $I$ denote voltage and current.

The analogy of acoustics, EM and IC systems in the series case and parallel case, and the corresponding realization approaches are summarized in Table \ref{tab:analogy}.
\begin{table}[]
\caption {Analogy of acoustic, EM and IC systems and the corresponding realization for series and parallel cases} \label{tab:analogy}
\begin{tabular}{cccc}
\hline\hline
                               & System type & Discontinuity & Realization                                                                             \\ \hline
\multirow{3}{*}{Series case}   & Acoustics  & $p$           & Membranes and plates   \\ 
                               & EM         & $\vv{E}$           & Impedance sheets that create magnetic current \\ 
                               & IC         & $V$             & Impedance in series                                                                     \\ \hline
\multirow{3}{*}{parallel case} & Acoustics  & $v$             & Side-loaded resonators     \\
                               & EM         & $-\hat{n}\times\vv{H}$          & Impedance sheets that create electric current  \\ 
                               & IC         & $I$             & Impedance in parallel                                                                   \\ \hline\hline
\end{tabular}
\end{table}

\section*{APPENDIX B: CONVERSION FROM TRANSFER MATRIX TO SCATTERING MATRIX}
	\setcounter{equation}{0}
	\renewcommand{\theequation}{B\arabic{equation}}
	\setcounter{figure}{0}
    \renewcommand{\thefigure}{B\arabic{figure}}

\begin{figure}
    \centering
    \includegraphics[width=0.6\linewidth]{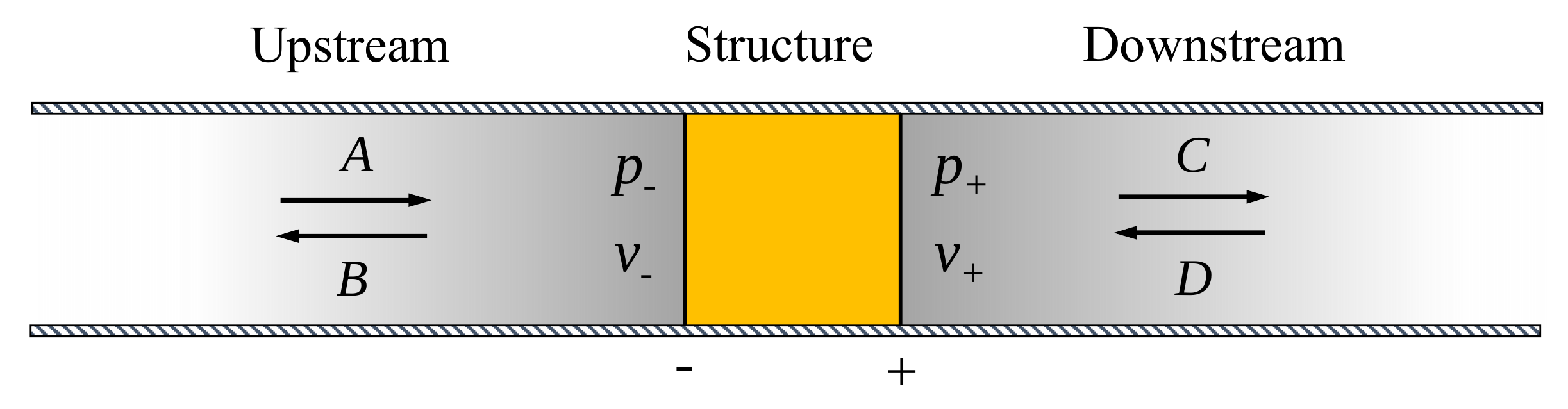}
    \caption{Schematic of the system for matrix conversion.}
    \label{fig:M_to_S}
\end{figure}

The schematic diagram is shown in Fig. \ref{fig:M_to_S}. The transfer matrix and scattering matrix of the structure are defined as
\begin{equation}
    \begin{bmatrix}
	\vdots\\
	p_-^{n-1}\\
	v_-^{n-1}\\
	p_-^{n}\\
	v_-^{n}\\
	p_-^{n+1}\\
	v_-^{n+1}\\
	\vdots
	\end{bmatrix}
	=
    M
	\begin{bmatrix}
	\vdots\\
	p_+^{n-1}\\
	v_+^{n-1}\\
	p_+^{n}\\
	v_+^{n}\\
	p_+^{n+1}\\
	v_+^{n+1}\\
	\vdots
	\end{bmatrix}
	\label{eq:B_PVM}
\end{equation}
\begin{equation}
    \begin{bmatrix}
	\vdots\\
	B^{n-1}\\
	C^{n-1}\\
	B^{n}\\
	C^{n}\\
	B^{n+1}\\
	C^{n+1}\\
	\vdots
	\end{bmatrix}
	=
    S
	\begin{bmatrix}
	\vdots\\
	A^{n-1}\\
	D^{n-1}\\
	A^{n}\\
	D^{n}\\
	A^{n+1}\\
	D^{n+1}\\
	\vdots
	\end{bmatrix}
	\label{eq:B_ABCDS}
\end{equation}
The calculation strategy of transfer matrix is given in Sec. II., and can be expressed as
\begin{equation}
    M
	=
	\begin{bmatrix}
	 & \vdots & \vdots & \vdots & \\
	\dots & M_{n-1,n-1}^{'} & M_{n-1,n}^{'} & M_{n-1,n+1}^{'} & \dots \\
	\dots & M_{n,n-1}^{'} & M_{n,n}^{'} & M_{n,n+1}^{'} & \dots \\
	\dots & M_{n+1,n-1}^{'} & M_{n+1,n}^{'} & M_{n+1,n+1}^{'} & \dots \\
	 & \vdots & \vdots & \vdots &  
	\end{bmatrix},
	\label{eq:B_M}
\end{equation}
where
\begin{equation}
    M_{i,j}^{'}=
    \begin{bmatrix}
    M_{2i-1,2j-1} & M_{2i-1,2j}\\
    M_{2i,2j-1} & M_{2i,2j}
    \end{bmatrix}
    \label{eq:B_M'}
\end{equation}
Using $A^n$, $B^n$, $C^n$ and $D^n$ to represent $p_-^n$, $v_-^n$, $p_+^n$ and $v_+^n$, we can rewrite Eq. \ref{eq:B_PVM} as
\begin{widetext}
\begin{equation}
    \begin{bmatrix}
	 & \vdots & \vdots & \vdots & \\
	\dots & S_{n-1,n-1}^{out} & S_{n-1,n}^{out} & S_{n-1,n+1}^{out} & \dots \\
	\dots & S_{n,n-1}^{out} & S_{n,n}^{out} & S_{n,n+1}^{out} & \dots \\
	\dots & S_{n+1,n-1}^{out} & S_{n+1,n}^{out} & S_{n+1,n+1}^{out} & \dots \\
	 & \vdots & \vdots & \vdots &  
	\end{bmatrix}
    \begin{bmatrix}
	\vdots\\
	Out^{n-1}\\
	Out^{n}\\
	Out^{n+1}\\
	\vdots
	\end{bmatrix}
	=
    \begin{bmatrix}
	 & \vdots & \vdots & \vdots & \\
	\dots & S_{n-1,n-1}^{in} & S_{n-1,n}^{in} & S_{n-1,n+1}^{in} & \dots \\
	\dots & S_{n,n-1}^{in} & S_{n,n}^{in} & S_{n,n+1}^{in} & \dots \\
	\dots & S_{n+1,n-1}^{in} & S_{n+1,n}^{in} & S_{n+1,n+1}^{in} & \dots \\
	 & \vdots & \vdots & \vdots &  
	\end{bmatrix}
	\begin{bmatrix}
	\vdots\\
	In^{n-1}\\
	In^{n}\\
	In^{n+1}\\
	\vdots
	\end{bmatrix},
	\label{eq:}
\end{equation}
\end{widetext}
where
\begin{equation}
    Out^n=\begin{bmatrix} B^n \\ C^n \end{bmatrix}
\end{equation}
\begin{equation}
    In^n=\begin{bmatrix} A^n \\ D^n \end{bmatrix}
\end{equation}
\begin{equation}
    S_{i,j}^{out}
    =
    \begin{bmatrix}
    \delta_{i j} & -\left ( M_{i,j}^{'}(1,1)+\dfrac{M_{i,j}^{'}(1,2)}{z_0} \right )\\
    -\dfrac{\delta_{i j}}{z_0} & -\left ( M_{i,j}^{'}(2,1)+\dfrac{M_{i,j}^{'}(2,2)}{z_0} \right)
    \end{bmatrix}
\end{equation}
\begin{equation}
    S_{i,j}^{in}
    =
    \begin{bmatrix}
    -\delta_{i j} & \left ( M_{i,j}^{'}(1,1)-\dfrac{M_{i,j}^{'}(1,2)}{z_0} \right )\\
    -\dfrac{\delta_{i j}}{z_0} & \left ( M_{i,j}^{'}(2,1)-\dfrac{M_{i,j}^{'}(2,2)}{z_0} \right )
    \end{bmatrix}
	\label{eq:}
\end{equation}
$\delta_{ij}$ is the Kronecker delta. Then, we can get the scattering matrix of the structure as
\begin{equation}
    S=\left [ S^{out} \right ]^{-1}S^{in}.
	\label{eq:}
\end{equation}

From the definition of the scattering matrix (Eq. \ref{eq:B_ABCDS}), we can see that the transmission and reflection coefficients for each order can be directly obtained from the corresponding elements in the scattering matrix. For simplicity, we represent the scattering matrix as

\begin{equation}
    S
	=
	\begin{bmatrix}
	 & \vdots & \vdots & \vdots & \\
	\dots & S_{n-1,n-1}^{'} & S_{n-1,n}^{'} & S_{n-1,n+1}^{'} & \dots \\
	\dots & S_{n,n-1}^{'} & S_{n,n}^{'} & S_{n,n+1}^{'} & \dots \\
	\dots & S_{n+1,n-1}^{'} & S_{n+1,n}^{'} & S_{n+1,n+1}^{'} & \dots \\
	 & \vdots & \vdots & \vdots &  
	\end{bmatrix},
	\label{eq:B_S}
\end{equation}
where
\begin{equation}
    S_{i,j}^{'}=
    \begin{bmatrix}
    r_{i,j}^+ & t_{i,j}^-\\
    t_{i,j}^+ & r_{i,j}^-
    \end{bmatrix}
    \label{eq:B_S'}
\end{equation}
with each element in $S_{i,j}^{'}$ represents the reflection and transmission coefficients for positive direction and negative direction, corresponding to the $i$th order output and $j$th order input.

\section*{APPENDIX C: CONVERGENCE OF THE RESULTS AS MORE ORDERS ARE TAKEN INTO ACCOUNT}
\setcounter{figure}{0}
\renewcommand{\thefigure}{C\arabic{figure}}

\begin{figure}
	\includegraphics[width=0.5\linewidth]{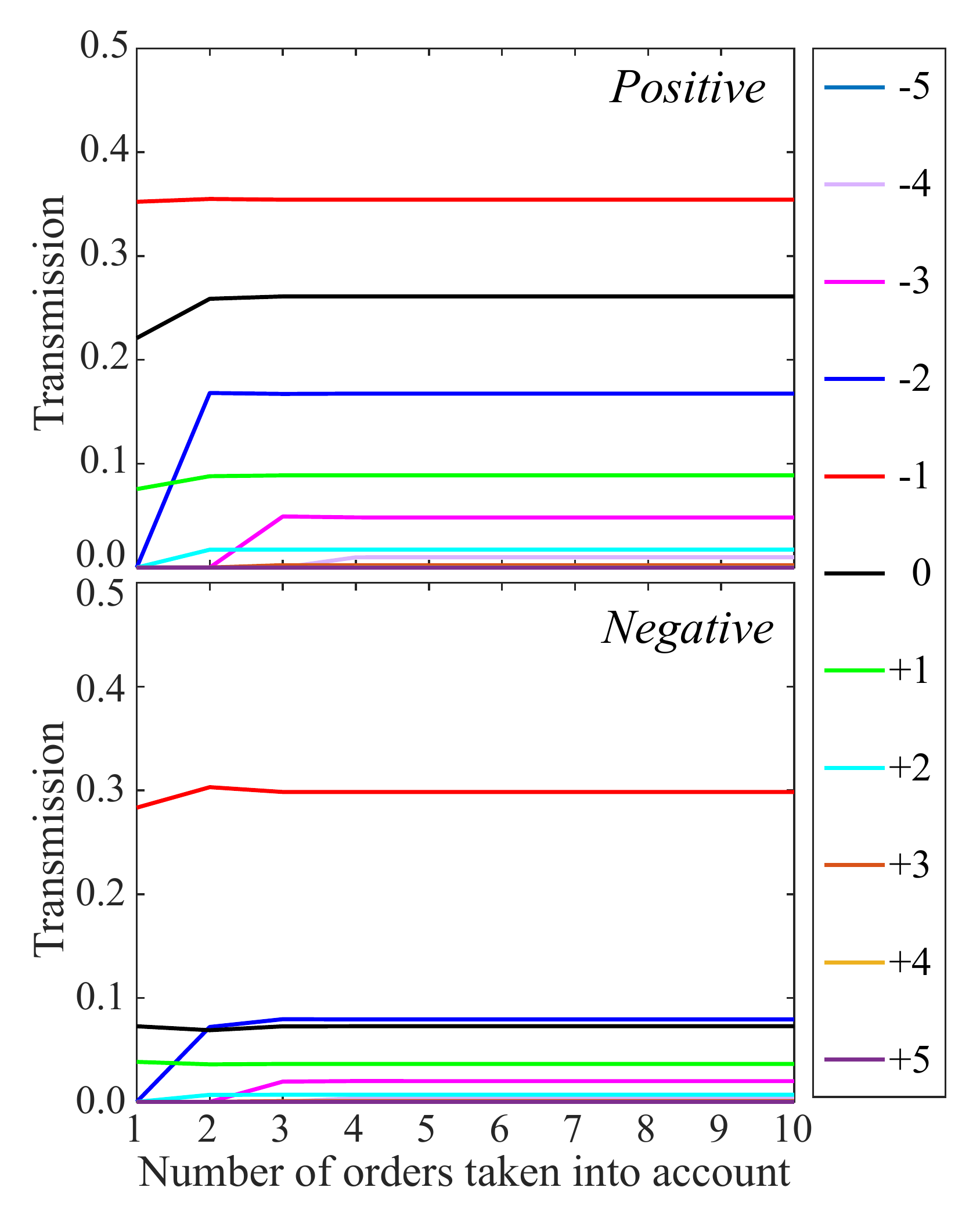}
	\caption{Transmission coefficients for 0 to $\pm 5$ order modes as we increase the number of orders taken into account. (a) and (b) showed the scenario when the incident wave is from positive direction and negative direction, respectively. The results remain unaffected after accounting for the $\pm5$ order.}  
	\label{fig:Number_of_orders}
\end{figure}

One of the main advantage for our proposed method is that it takes high order waves into account. Then it comes to the question of what kind of truncation gives reliable results. Here we pick the 4-resonator non-reciprocal sound transmission case as an example. Fig. \ref{fig:Number_of_orders} shows the convergence of transmission coefficient amplitudes for 0 to $\pm 5$ order waves as we increase the number of modes taken into account during the truncation process. Here the transmission coefficient for each order is defined as the ratio between the complex amplitude for each order and the incident wave amplitude. They can be directly obtained from the corresponding components in the scattering matrix. From Fig. \ref{fig:Number_of_orders} we can see that high order harmonics do play a role, and sometimes they are higher than the $0$th order, therefore, omitting them during calculation will lead to inaccurate results. In our example case, the amplitudes of harmonics higher than $\pm 5$th order are lower than $0.01$, and therefore can be neglected. The amplitudes of 0 to $\pm 5$ orders remain essentially unchanged after the $5$th order is taken into account. This justifies our calculation where we truncated the matrix at $\pm 10$ order.
 
\bibliographystyle{apsrev4-1}
\bibliography{references}

\end{document}